\documentclass[conference]{IEEEtran}
\IEEEoverridecommandlockouts
\usepackage{amsmath,amsfonts,amssymb}
\usepackage{graphicx}
\usepackage{cite}
\usepackage{booktabs}
\usepackage{algorithm}
\usepackage{algorithmic}
\usepackage{microtype}
\usepackage{color}

\newcommand{\Nset}{\mathcal N}
\newcommand{\Bset}{\mathcal B}

\begin{document}

\title{Timely Ground-Station Status Delivery via UAV-Assisted Data Collection and Hierarchical RL%
\thanks{Author's accepted manuscript.
Accepted for publication in the proceedings of the
2026 Asilomar Conference on Signals, Systems, and Computers.}
}

\author{
\IEEEauthorblockN{Xiangjian Gao\textsuperscript{1}, Mohammad~Moltafet\textsuperscript{2}, and Hamid R. Sadjadpour\textsuperscript{1}}
\IEEEauthorblockA{\textsuperscript{1}Department of Electrical and Computer Engineering, University of California Santa Cruz, CA, USA\\
\textsuperscript{2}Department of Electrical and Computer Engineering, Tennessee Tech University, TN, USA\\
Email: xgao41@ucsc.edu; mmoltafet@tntech.edu; hamid@ucsc.edu}}

\maketitle

\begin{abstract}
We consider a UAV-assisted status-update system in which a rotary-wing unmanned aerial vehicle (UAV) collects updates from multiple sensors, stores the newest update of each sensor, and forwards selected buffered updates to a ground station (GS). We minimize a weighted sum of the time-average GS-side age of information (AoI) and UAV energy consumption over a finite horizon. The optimization is subject to half-duplex operation, sensor and UAV energy-causality constraints, transmission-rate and power limits, and a terminal-location constraint. Because collection changes the information age at the UAV but does not reduce the GS-side AoI until the buffered update is forwarded, mobility, collection, and forwarding decisions are coupled over time. We use a hierarchical AoI--energy actor--critic (HAE-AC) controller in which a coarse-timescale actor selects return-feasible UAV movements, while a deterministic per-slot scheduler constructs feasible collection or forwarding sets and assigns the minimum transmit power required for each scheduled transmission. In a 100-sensor deployment, HAE-AC reduces at the GS by about 22\% relative to an adaptive staleness-gated benchmark. Compared with forward-upon-collection, selective buffered forwarding reduces the number of downlink bursts by roughly 30\% while increasing the number of collections per sensor by more than 23\%.
\end{abstract}

\begin{IEEEkeywords}
Age of information, UAV communications, trajectory design, scheduling, actor--critic methods.
\end{IEEEkeywords}

\section{Introduction}
Age of information (AoI) measures the time elapsed since the generation of the newest status update available at a destination~\cite{kaul2012status,yates2021survey}. UAV mobility extends sensing coverage to geographically distributed or low-power sensors~\cite{liu2021freshness} and to infrastructure-limited settings such as post-disaster search and rescue~\cite{gao2026postdisaster}, but it also couples communication scheduling with trajectory design. Over a finite mission, the UAV must therefore decide where to move, which updates to collect and forward, and how much energy to spend while meeting its terminal-location constraint.

For a relay system, the AoI at the UAV and the AoI at the final destination are different states. A newly collected update refreshes the copy stored at the UAV, but the GS-side AoI continues to increase until that update is forwarded. Previous UAV-AoI studies have mainly considered collection-side freshness through joint trajectory and scheduling design~\cite{liu2021freshness,wu2022aiframework,zhu2023transformer,gao2023multiuav,huang2025iterative}. Recent work has also coupled uplink acquisition and downlink delivery~\cite{zhao2025updownlink}. Learning-based scheduling for multi-source relaying has been studied without UAV mobility~\cite{zakeri2024relay}. In the system considered here, the UAV stores the newest update of each sensor and the difference between UAV-side and GS-side AoI determines whether a buffered update can improve the information at the GS.

Forwarding also competes with collection because the UAV is half duplex. Sending a downlink burst consumes one communication slot and incurs a fixed burst overhead and circuit activation cost. Forwarding after every collection therefore uses more downlink slots, whereas waiting too long increases the age of buffered updates. We allow several useful buffered updates to share one downlink burst and make the forwarding time part of the control decision.

Hierarchical and two-timescale learning have been used to separate UAV mobility from communication scheduling in related trajectory-control problems~\cite{ferdowsi2021neural,gong2023hierarchical,fu2026twolayer,tian2024twotimescale}; the two-layer design of~\cite{fu2026twolayer} likewise pairs a learned trajectory policy with a non-learned scheduler. Our hierarchical AoI--energy actor--critic (HAE-AC) controller follows this separation, but its scheduling subproblem also selects the forwarding time and the subset of buffered updates that share a burst, whose required transmit power is not additive across the selected updates. A deterministic per-slot scheduler handles collection, forwarding, and transmit power under the communication and energy constraints, so the movement actor operates on a coarser timescale without the combinatorial sensor-selection decisions. We compare HAE-AC with a single-layer actor--critic and two movement heuristics, and evaluate forward-upon-collection (FUC), Selective, and Threshold forwarding. Full details and extended results are given in the journal version~\cite{gao2026journal}.

\section{System Model}
\looseness=-1
We consider $N$ static sensors, $\Nset=\{1,\ldots,N\}$, one rotary-wing UAV, and one GS. Time is divided into $T$ slots of duration $\delta$ indexed by $t$, where each slot corresponds to the interval $[t\delta,(t+1)\delta)$. The UAV flies at a fixed altitude $z$ above the ground plane, so $\mathbf L(t)=(x(t),y(t),z)$, while sensor $i$ and the GS lie on the ground plane at $\mathbf L_i=(x_i,y_i,z_i)$ and $\mathbf L_{\rm G}=(x_{\rm G},y_{\rm G},z_{\rm G})$; the norms below are three-dimensional distances. In each slot the UAV selects a horizontal movement $\mathbf m(t)\in\mathcal M\triangleq\{\mathbf e_{\rm N},\mathbf e_{\rm S},\mathbf e_{\rm E},\mathbf e_{\rm W},\mathbf e_{\rm H}\}$, where $\mathbf e_{\rm N}=(0,1,0)$, $\mathbf e_{\rm S}=(0,-1,0)$, $\mathbf e_{\rm E}=(1,0,0)$, and $\mathbf e_{\rm W}=(-1,0,0)$ are the unit direction vectors toward north, south, east, and west, and $\mathbf e_{\rm H}=(0,0,0)$ denotes hovering. Its position evolves as
\begin{equation}
\mathbf L(t+1)=\mathbf L(t)+\delta v_{\mathbf m(t)}\mathbf m(t),\quad t=0,\ldots,T-2,
\label{eq:mobility}
\end{equation}
where $v_{\mathbf m(t)}=v$ for $\mathbf m(t)\neq\mathbf e_{\rm H}$ and $0$ for hover. The initial and terminal locations satisfy $\mathbf L(0)=\mathbf L_{\rm I}$ and $\mathbf L(T-1)=\mathbf L_{\rm F}$; the UAV remains at the terminal location through the final slot. The UAV can communicate while moving and stays inside a flight region $\mathcal L_{\rm fly}$; the GS lies outside it. The large-scale channel gain in slot $t$ is evaluated at $\mathbf L(t)$ and is treated as constant during that slot.

Let $I^{\rm U}(t)$ and $I^{\rm D}(t)$ indicate uplink collection and downlink forwarding. Half-duplex operation requires
\begin{equation}
I^{\rm U}(t)+I^{\rm D}(t)\le 1.
\label{eq:halfduplex}
\end{equation}
Sensor $i$ is scheduled by $b_i(t)\in\{0,1\}$, with $b_i(t)\le I^{\rm U}(t)$ and $\sum_i b_i(t)\le\beta$; the scheduled sensors transmit simultaneously over orthogonal uplink channels, each of bandwidth $W^{\rm U}$, so that the UAV collects up to $\beta$ updates in one slot. The uplink channel gain and achievable bits per slot are
\begin{align}
h_i^{\rm u}(t)&=h_0^{\rm U}\|\mathbf L(t)-\mathbf L_i\|^{-\alpha^{\rm U}},\\
R_i^{\rm u}(t)&=\delta W^{\rm U}\log_2\!\left(1+\frac{P_i^{\rm u}(t)h_i^{\rm u}(t)}{N_0W^{\rm U}}\right).
\end{align}
Here $h_0^{\rm U}$ is the reference gain at 1~m, $\alpha^{\rm U}$ the path-loss exponent, $W^{\rm U}$ the uplink bandwidth, $P_i^{\rm u}(t)$ the sensor transmit power, and $N_0$ the noise power spectral density; the downlink uses $h_0^{\rm D}$, $\alpha^{\rm D}$, and $W^{\rm D}$. Sensor energy evolves as $E_i(t+1)=E_i(t)-P_i^{\rm u}(t)\delta$ and must remain nonnegative.

The UAV keeps only the newest collected update of each sensor. Let $q_i(t)\in\{0,1\}$ indicate that sensor $i$ is included in a downlink burst. Conditional on $I^{\rm D}(t)=1$, the burst size is
\begin{equation}
S^{\rm D}(t)=S^{\rm H}+\sum_i q_i(t)S_i,
\label{eq:payload}
\end{equation}
where $S_i$ is the size of sensor $i$'s update and $S^{\rm H}$ is the fixed burst overhead. With $h_{\rm u}^{\rm G}(t)=h_0^{\rm D}\|\mathbf L(t)-\mathbf L_{\rm G}\|^{-\alpha^{\rm D}}$, the downlink supports
\begin{equation}
R^{\rm D}(t)=\delta W^{\rm D}\log_2\!\left(1+\frac{P_{\rm u}^{\rm G}(t)h_{\rm u}^{\rm G}(t)}{N_0W^{\rm D}}\right),
\end{equation}
and a scheduled burst must satisfy $I^{\rm D}(t)S^{\rm D}(t)\le R^{\rm D}(t)$. Uplink updates and downlink bursts are assumed to complete within one slot, with no retransmission or partial reception; under unreliable links this understates the delivery time. Sensor idle/listening energy and UAV reception energy are neglected relative to active transmission and UAV propulsion.

Let $\Delta_i^{\rm U}(t)$ and $\Delta_i^{\rm G}(t)$ denote the AoI of sensor $i$ at the UAV and GS, both initialized to $1$. Their slot-to-slot evolution is
\begin{align}
\Delta_i^{\rm U}(t+1)&=1+\big(1-b_i(t)\big)\Delta_i^{\rm U}(t),\label{eq:aoiu}\\
\Delta_i^{\rm G}(t+1)&=1+q_i(t)\Delta_i^{\rm U}(t)+\big(1-q_i(t)\big)\Delta_i^{\rm G}(t).\label{eq:aoig}
\end{align}
For $q_i(t)=1$, \eqref{eq:aoig} uses the newest buffered copy at the UAV; the additional slot accounts for the downlink transmission.

\subsection{Problem Formulation}
For priority weights $w_i\ge0$ with $\sum_iw_i=1$, the weighted time-average GS-side AoI is
\begin{equation}
\bar\Delta^{\rm G}=\frac{1}{T}\sum_{t=0}^{T-1}\sum_iw_i\Delta_i^{\rm G}(t).
\end{equation}
The UAV slot energy is $E_{\rm slot}(t)=E_{\rm fly}(t)+I^{\rm D}(t)(P_{\rm u}^{\rm G}(t)+P_{\rm c})\delta$, where $P_{\rm c}$ is the transmit-chain circuit power drawn during a downlink burst and $E_{\rm fly}(t)$ follows the rotary-wing propulsion model in~\cite{zeng2019rotary}; related UAV AoI--energy formulations are considered in~\cite{zhang2024aoienergy}. The UAV energy state evolves as $E^{\rm U}(t+1)=E^{\rm U}(t)-E_{\rm slot}(t)$ and must remain nonnegative. Define
\begin{equation}
J=\bar\Delta^{\rm G}+\lambda\sum_{t=0}^{T-1}E_{\rm slot}(t),
\label{eq:objective}
\end{equation}
where $\lambda\ge0$ weights UAV energy against GS-side freshness. With
$\mathcal X=\{\mathbf m,P_i^{\rm u},P_{\rm u}^{\rm G},b_i,q_i,I^{\rm U},I^{\rm D}\}_{t,i}$,
the finite-horizon problem can be written compactly as
\begin{subequations}
\label{eq:P0}
\begin{align}
\text{(P0)}\quad \min_{\mathcal X}\quad &J, \label{eq:P0_obj}\\
\text{s.t.}\quad
&\mathbf L(0)=\mathbf L_{\rm I},\quad \mathbf L(T-1)=\mathbf L_{\rm F}, \label{eq:P0_terminal}\\
&\mathbf L(t)\in\mathcal L_{\rm fly},\quad
I^{\rm U}(t)+I^{\rm D}(t)\le1, \label{eq:P0_movehd}\\
&b_i(t)\le I^{\rm U}(t),\; q_i(t)\le I^{\rm D}(t),\;
\sum_i b_i(t)\le\beta, \label{eq:P0_mode}\\
&b_i(t)S_i\le R_i^{\rm u}(t),\;
I^{\rm D}(t)S^{\rm D}(t)\le R^{\rm D}(t), \label{eq:P0_rate}\\
&0\le P_i^{\rm u}(t)\le b_i(t)P_i^{{\rm u},\max}, \label{eq:P0_power_u}\\
&0\le P_{\rm u}^{\rm G}(t)\le I^{\rm D}(t)P_{\rm u}^{{\rm G},\max}, \label{eq:P0_power_d}\\
&E_i(t)\ge0,\quad E^{\rm U}(t)\ge0. \label{eq:P0_energy}
\end{align}
\end{subequations}
These constraints are imposed together with the state and buffer dynamics described above. The binary mode and scheduling variables, position-dependent rates, and coupling of current actions to future position, energy, buffer contents, and GS-side AoI make \eqref{eq:P0} a mixed-integer nonconvex problem with decisions coupled across time. We therefore formulate it as a finite-horizon MDP.

\section{HAE-AC Controller}
\subsection{MDP}
The slot-start state is
\begin{equation}
s_t=\big(t,\mathbf L(t),E^{\rm U}(t),(E_i(t),\Delta_i^{\rm U}(t),\Delta_i^{\rm G}(t))_{i\in\Nset}\big).
\label{eq:state}
\end{equation}
A buffered update is useful for the GS only when it is newer than the GS copy. We define the pending set
\begin{equation}
\Bset(t)=\{i\in\Nset:\Delta_i^{\rm U}(t)<\Delta_i^{\rm G}(t)\}.
\label{eq:pending}
\end{equation}
The full action $a_t$ contains movement, communication mode, and the collection and forwarding sets. Transmit powers are determined after a schedule is selected. For a scheduled uplink,
\begin{equation}
\widehat P_i^{\rm u}(t)=\frac{N_0W^{\rm U}}{h_i^{\rm u}(t)}\left(2^{S_i/(\delta W^{\rm U})}-1\right),
\label{eq:pminu}
\end{equation}
and for a downlink burst of size $S^{\rm D}(t)$,
\begin{equation}
\widehat P_{\rm u}^{\rm G}(t)=\frac{N_0W^{\rm D}}{h_{\rm u}^{\rm G}(t)}\left(2^{S^{\rm D}(t)/(\delta W^{\rm D})}-1\right).
\label{eq:pmind}
\end{equation}
These are the minimum powers that satisfy the one-slot rate constraints; unscheduled links use zero transmit power.

\looseness=-1
The per-slot cost is $C(s_t,a_t)=T^{-1}\sum_iw_i\Delta_i^{\rm G}(t)+\lambda E_{\rm slot}(t)$. The movement actor is trained with the shaped reward
\begin{equation}
r_t=-C(s_t,a_t)+\chi\widetilde{\Phi}(s_{t+1})-\widetilde{\Phi}(s_t),
\label{eq:reward}
\end{equation}
where $\chi<1$ is the training discount and
\begin{equation}
\widetilde{\Phi}(s_t)=-\frac{1}{T}\sum_iw_i\big(\Delta_i^{\rm G}(t)+\kappa\Delta_i^{\rm U}(t)\big)-c_{\rm nav}\Psi_{\rm nav}(s_t).
\label{eq:trainingpotential}
\end{equation}
\looseness=-1
Here $\Psi_{\rm nav}(s_t)$ is the normalized soft-min distance from the UAV to the high-staleness sensors $\mathcal N_{\rm st}(t)=\{i\in\Nset:w_i\Delta_i^{\rm U}(t)\ge\tfrac12\max_{j\in\Nset}w_j\Delta_j^{\rm U}(t)\}$, so the term $-c_{\rm nav}\Psi_{\rm nav}$ increases as the UAV moves closer to them. The terminal potential is set to zero. Shaping and discounting are used only for training; model selection and all reported numerical results use the undiscounted objective in \eqref{eq:objective}.

\begin{figure}[t]
\centering
\includegraphics[width=0.82\columnwidth]{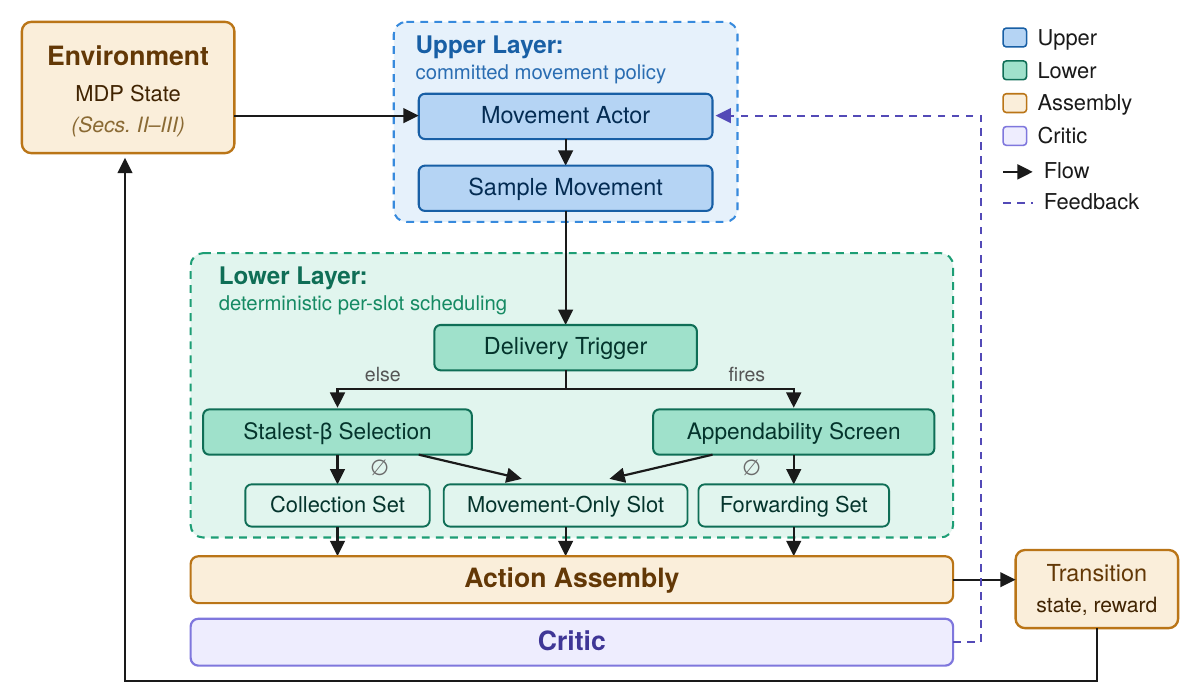}
\caption{HAE-AC architecture: a coarse-timescale movement actor and a deterministic per-slot scheduler, with fallback from forwarding to collection when no feasible burst is formed.}
\label{fig:architecture}
\vspace{-1.5mm}
\end{figure}

\subsection{Control Structure}
\looseness=-1
A single-layer actor--critic (SL-AC) that learns movement and collection scheduling jointly faces, at each collection opportunity, a movement--collection action class of size at most
\begin{equation}
|\mathcal M_t|\sum_{k=0}^{\beta}{N\choose k}\le5\sum_{k=0}^{\beta}{N\choose k},
\label{eq:slacspace}
\end{equation}
\looseness=-1
where $\mathcal M_t$ is the admissible movement set. SL-AC does not learn forwarding, and its unmasked collection heads are projected after sampling by feasibility filtering and top-$\beta$ truncation. HAE-AC instead learns only the movement decision (Fig.~\ref{fig:architecture}). A candidate movement is admissible, i.e., belongs to $\mathcal M_t$, if its next position stays in the flight region, the remaining movement transitions can still reach $\mathbf L_{\rm F}$ in axis-aligned steps, and the remaining UAV energy covers the terminal reserve, namely the propulsion energy of a shortest axis-aligned return path from the candidate's next position to $\mathbf L_{\rm F}$ plus the hovering energy for the slots remaining thereafter. Each admissible candidate $\mathbf m\in\mathcal M_t$ is represented by a fixed-dimensional feature vector $\mathbf z_t(\mathbf m)$ that collects the slot index, UAV energy, distances to the GS and to $\mathbf L_{\rm F}$, weighted-AoI, sensor-energy, and pending-buffer summaries, and a directional staleness aggregate $\sum_i w_i\Delta_i^{\rm U}(t)$ over the sensors in the $90^\circ$ cone centered on $\mathbf m$ (for hover, over the sensors collectable at $\mathbf L(t)$); the actor scores each $\mathbf z_t(\mathbf m)$ and samples the movement from a softmax over the at most five admissible scores. The complete definition of $\mathbf z_t(\mathbf m)$ and the actor--critic network architecture are given in~\cite{gao2026journal}. The selected movement is held for up to $\mu$ slots and is resampled earlier if it becomes inadmissible.

\looseness=-1
The per-slot scheduler first evaluates the selected forwarding rule. If forwarding is not attempted, it chooses up to $\beta$ feasible sensors with the largest $w_i\Delta_i^{\rm U}(t)$ among the eligible ones, i.e., those with $\Delta_i^{\rm U}(t)\ge2$. If forwarding is attempted, only sensors in $\Bset(t)$ are considered. The forwarding rules use fixed state-based triggers and stopping conditions that do not read $\lambda$ directly, and pending candidates are examined in descending order of $w_i(\Delta_i^{\rm G}-\Delta_i^{\rm U})$.  If a forwarding attempt produces an empty feasible set, the scheduler falls back to collection.

\subsection{Forwarding Rules and Training}
\looseness=-1
The three evaluated forwarding rules share the same pending-set ordering, appendability tests, minimum-power assignment, and fallback-to-collection mechanism, and differ only in when a burst is attempted. FUC attempts a downlink whenever $\Bset(t)$ is nonempty; Selective and Threshold wait until the pending set or a top-ranked pending gap $\Delta_i^{\rm G}-\Delta_i^{\rm U}$ exceeds a threshold and examine a bounded number of candidates; Selective also forwards in the final slots and withholds bursts that would carry too few updates. The constants are given in Section~\ref{sec:results}; the appendability tests, stopping conditions, and the fallback mechanism are described in detail in~\cite{gao2026journal}.

The movement actor is trained on-policy with PPO~\cite{schulman2017ppo}, using a state-value critic for generalized advantage estimation. A movement commitment of $\tau$ slots is treated as one transition with a $\chi^\tau$ bootstrap. The per-slot scheduler is deterministic and contributes no policy log-probability.

\begin{algorithm}[t]
\caption{Per-slot HAE-AC control}
\label{alg:haeac}
\begin{algorithmic}[1]
\STATE Observe $s_t$ and update the return-feasible movement set.
\IF{a new movement decision is required}
  \STATE Select $\mathbf m(t)$ with the movement actor.
\ENDIF
\STATE Evaluate uplink and downlink feasibility at $\mathbf L(t)$; set $\mathcal F\leftarrow\emptyset$.
\IF{the forwarding rule triggers and $\Bset(t)\neq\emptyset$}
  \STATE Construct a feasible forwarding set $\mathcal F\subseteq\Bset(t)$.
\ENDIF
\IF{$\mathcal F\neq\emptyset$}
  \STATE Use \eqref{eq:pmind} and forward $\mathcal F$.
\ELSE
  \STATE Select up to $\beta$ eligible-feasible sensors and use \eqref{eq:pminu}.
\ENDIF
\STATE Execute the slot and update position, buffer, AoI, and energy states.
\end{algorithmic}
\end{algorithm}

\subsection{Feasibility and Complexity}
\looseness=-1
The movement screening preserves the flight region and the time and propulsion reserve needed for the terminal-location constraint. Uplink candidates are retained only if their eligibility, rate, peak-power, sensor-energy, and cardinality constraints are satisfied. Downlink candidates are sorted once and scanned once; each candidate is admitted if the enlarged burst remains power- and UAV-energy-feasible and is skipped otherwise, so the scan continues past infeasible candidates, and the fallback collection rule is applied before the realized communication mode is assigned. These checks also enforce half-duplex operation, so the assembled per-slot action satisfies the corresponding constraints in \eqref{eq:P0}. Algorithm~\ref{alg:haeac} summarizes the resulting per-slot control. The online complexity is $O(d_sd_h+\beta N+|\Bset(t)|\log|\Bset(t)|)$, where $d_s$ and $d_h$ are the movement-actor feature dimension and hidden width.

\section{Numerical Results}\label{sec:results}
We evaluate $N=100$ sensors uniformly distributed over a $2000\times2000$~m field with one-second slots and at most $\beta=8$ collections per slot; the GS is outside the monitored field and the UAV starts and ends at a corner depot. Table~\ref{tab:params} lists the remaining settings. Selective triggers at $|\Bset(t)|\ge16$, at a gap of 28 slots among the four highest-ranked candidates, or at $t\ge T-6$, examines at most 20 candidates, and outside the final six slots withholds a burst when three or more updates are pending but fewer than three fit; Threshold triggers at $|\Bset(t)|\ge18$ or when the largest pending gap reaches 34 slots, with at most 12 candidates. These constants were selected on validation instances and then fixed for all controllers, test instances, and $\lambda$ values. Training uses the curriculum $(N,\text{field},T)=(25,1000\,\mathrm m,300)\rightarrow(50,1400\,\mathrm m,600)\rightarrow(100,2000\,\mathrm m,1200)$, with $\kappa=0.5$, $\chi=0.997$, PPO clip $0.2$, GAE parameter $0.95$, and learning rate $3\times10^{-4}$. Further implementation details, including the network sizes and the validation protocol, are given in~\cite{gao2026journal}.

Learned methods use three training seeds; checkpoint selection uses eight validation instances and reported results use 16 disjoint test instances. Coverage denotes the fraction of sensors collected at least once during the mission. HAE-AC is compared with a distance-greedy patroller, an adaptive staleness-gated heuristic (ASG) that steers toward the nearest sensor whose UAV-side age exceeds the network mean, and SL-AC. ASG uses the same return screening, collection rule, forwarding construction, and power assignment as HAE-AC, so their FUC comparison isolates the movement rule.

\begin{table}[t]
\centering
\caption{Main simulation settings.}
\label{tab:params}
\scriptsize
\setlength{\tabcolsep}{4.0pt}
\renewcommand{\arraystretch}{1.04}
\begin{tabular}{ll}
\toprule
Parameter & Value\\
\midrule
Sensors / field / horizon $T$ & $100$ / $2000\times2000$ m / $1200$ slots ($\delta=1$ s)\\
Altitude / speed / $\beta$ & $50$ m / $15$ m/s / $8$\\
Packet sizes $S_i$ / overhead $S^{\rm H}$ & $\{2048,4096,8192\}$ / $2048$ bits\\
Sensor peak power / battery & $\{0.05,0.1,0.2,0.5\}$ $\mu$W / $\mathcal U[1.5,8]$ J\\
UAV downlink peak / circuit power $P_{\rm c}$ & $0.2$ W / $3$ W\\
Uplink $(W^{\rm U},f_c,\alpha^{\rm U})$ & $125$ kHz / $915$ MHz / $2.6$\\
Downlink $(W^{\rm D},f_c,\alpha^{\rm D})$ & $180$ kHz / $2.1$ GHz / $2.4$\\
Noise PSD $N_0$ & $-174$ dBm/Hz\\
UAV energy / $\lambda$ / $\mu$ & $222.4$ kJ / $10^{-3}$ / $10$\\
\bottomrule
\end{tabular}
\vspace{-1mm}
\end{table}

\begin{figure}[t]
\centering
\includegraphics[width=0.82\columnwidth]{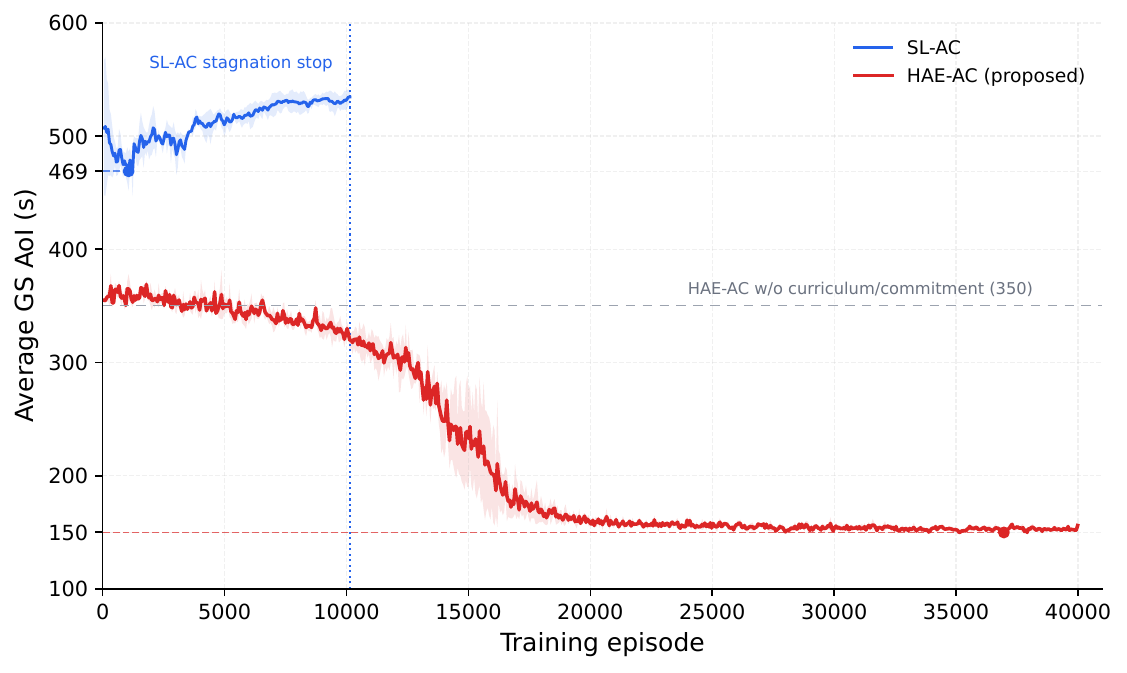}
\caption{Full-scale training histories at $N=100$ for HAE-AC, SL-AC, and the no-curriculum, no-commitment HAE-AC ablation.}
\label{fig:conv}
\vspace{-1.5mm}
\end{figure}

\begin{figure}[t]
\centering
\includegraphics[width=0.94\columnwidth]{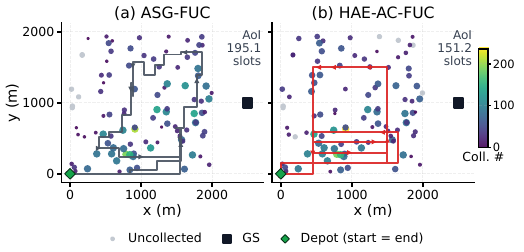}
\caption{UAV trajectories on one held-out deployment under FUC: (a) ASG and (b) HAE-AC, with the same lower-layer rules; AoI are test-set averages.}
\label{fig:trajectory}
\vspace{-1.5mm}
\end{figure}

\looseness=-1
Under FUC, the distance-greedy patroller gives 534.0~slots average GS-side AoI and 11.4\% collection coverage. ASG gives 195.1~slots. HAE-AC gives $151.2\pm2.0$~slots with $93.6\pm0.9\%$ coverage, corresponding to a 22.5\% reduction relative to ASG. SL-AC gives $454.6\pm21.9$~slots with $54.1\pm8.1\%$ coverage. Fig.~\ref{fig:conv} shows the corresponding training behavior, and Fig.~\ref{fig:trajectory} the path geometry on one held-out deployment, where ASG follows its staleness-gated steering rule while HAE-AC patrols the field more widely.

\begin{table}[t]
\centering
\caption{Architecture and forwarding-rule comparison. Learned rows report the mean and sample standard deviation over three training seeds. Coll./s.\ is the number of collections per sensor.}
\label{tab:main}
\scriptsize
\setlength{\tabcolsep}{1.55pt}
\renewcommand{\arraystretch}{1.05}
\begin{tabular}{llccccc}
\toprule
Rule & Controller & AoI (slots) & Cov. (\%) & Coll./s. & Bursts & kJ\\
\midrule
FUC & HAE-AC & $151.2\pm2.0$ & $93.6\pm0.9$ & 47.7 & 599 & 168.1\\
    & ASG    & 195.1 & 93.3 & 47.7 & 599 & 168.1\\
    & SL-AC  & $454.6\pm21.9$ & $54.1\pm8.1$ & 42.9 & 600 & 178.3\\
\midrule
Selective & HAE-AC & $147.7\pm0.8$ & $94.5\pm0.8$ & 58.9 & 421 & 167.6\\
          & ASG    & 197.2 & 93.3 & 60.1 & 355 & 167.4\\
\midrule
Threshold & HAE-AC & $149.5\pm3.4$ & $93.9\pm1.2$ & 61.8 & 378 & 167.5\\
          & ASG    & 194.6 & 94.4 & 63.6 & 315 & 167.3\\
\bottomrule
\end{tabular}
\vspace{-1mm}
\end{table}

\looseness=-1
Table~\ref{tab:main} separates the movement comparison from the forwarding comparison. For HAE-AC, changing FUC to Selective reduces the burst count from 599 to 421 and increases collections per sensor from 47.7 to 58.9. The average GS-side AoI changes from 151.2~slots to 147.7~slots, and Threshold gives HAE-AC its fewest bursts (378) at $149.5\pm3.4$~slots. The ASG rows isolate the forwarding-rule effect from the movement policy.

\subsection{Sensitivity and Scaling}
As an energy-causality diagnostic, we extend the horizon to $T=3600$ slots with the same $222.4$-kJ UAV budget and no retraining. The controllers return and land on average between slots 1588 and 1596; during powered operation, HAE-AC maintains 165--176~slots GS-side AoI versus 212--218~slots for ASG. For this diagnostic, the reserve includes return-flight energy but omits post-return hovering.

\looseness=-1
For the Selective controller, raising $\lambda$ from $0$ to $10^{-3}$ (one independently trained seed per $\lambda$ value, hence the small difference from the three-seed mean in Table~\ref{tab:main}) changes the average AoI from 149.1~slots to 148.0~slots and leaves the UAV energy at about 167.6~kJ. Intermediate probes place the transition between $\lambda=10^{-2}$ and $10^{-1}$: the corresponding AoI values are 150.1~slots and 414.9~slots. At $\lambda=10$, the UAV uses 166.6~kJ, sends about 107 bursts, and the AoI is 473.6~slots; the $\lambda=1$ and $\lambda=10$ operating points are nearly identical. Propulsion therefore dominates the total energy budget, and large $\lambda$ mainly trades freshness and coverage for a small change in total UAV energy.

At similar sensor density, $N\in\{25,50,100,150\}$, HAE-AC under Selective obtains average GS-side AoI of $31.0$, $70.4$, $147.7$, and $258.9$ slots versus $30.6$, $84.3$, $197.2$, and $301.2$ slots for ASG-Selective: the two are comparable at $N=25$ and HAE-AC is lower at the three larger scales. SL-AC under FUC rises from $78.1$ to $650.0$ slots over the same range. In a separate priority sweep using the same validation-selected Selective checkpoint without retraining, increasing the relative weight of a randomly selected 20\% sensor group from $w=1$ to $w=16$ changes that group's mean AoI from 146.0~slots to 135.6~slots, while the mean AoI of the remaining sensors changes from 148.6~slots to 163.1~slots. The weights are renormalized to sum to one for each test.

\subsection{Clustered Deployment}
We also test the three validation-selected Selective HAE-AC checkpoints without retraining on a different mission setting: $N=100$ sensors in five spatial clusters, a $T=600$-slot horizon, and a $130$-kJ UAV budget. HAE-AC obtains $96.6\pm1.4$~slots GS-side AoI with $93.4\%$ coverage, compared with $125.3$~slots and $86.9\%$ for the forwarding-rule-matched ASG-Selective reference. Since the geometry, horizon, and energy budget change together, this is a combined robustness test rather than a test of spatial generalization alone.

\section{Conclusion}
\looseness=-1
We considered a UAV relay that stores sensor updates and delivers selected buffered updates to a GS. The formulation tracks UAV-side and GS-side AoI separately under half-duplex communication, energy, and terminal-location constraints. HAE-AC learns only UAV movement and constructs communication schedules and minimum-power assignments deterministically. At $N=100$ under FUC, HAE-AC reduces average GS-side AoI from 195.1~slots for ASG to 151.2~slots. With HAE-AC, Selective forwarding reduces the burst count from 599 to 421 and increases collections per sensor from 47.7 to 58.9. Network-size and priority tests characterize scaling and weight adaptation, while the clustered-deployment test preserves the HAE-AC--ASG AoI gap without retraining under a joint change in mission conditions.

\end{document}